\documentclass[conf]{new-aiaa}
\usepackage[utf8]{inputenc}

\usepackage{graphicx}
\usepackage{caption}
\usepackage{subcaption}
\usepackage{amsmath}
\DeclareMathOperator*{\argmax}{argmax}
\usepackage[version=4]{mhchem}
\usepackage{siunitx}
\usepackage{longtable,tabularx}
\usepackage{multirow}
\usepackage{algorithm}
\usepackage{algpseudocode}
\usepackage{todonotes}
\usepackage{fancyhdr}

\fancypagestyle{FirstPage}{
\lfoot{\textit{\small Copyright © 2024 by Delft University of Technology, Faculty of Aerospace Engineering.\\ Preprint submitted to AIAA SciTech 2024 Forum.}} 
}

\usepackage[style=authoryear,maxbibnames=9,maxcitenames=2,backend=biber]{biblatex}
\title{High-Dimensional Bayesian Optimisation with Large-Scale Constraints - An Application to Aeroelastic Tailoring}

\author{
	Hauke Maathuis\footnote{Ph.D. Candidate, Aerospace Structures and Materials, Faculty of Aerospace Engineering, Delft University of Technology, h.f.maathuis@tudelft.nl},
    Roeland De Breuker \footnote{Associate Professor, Aerospace Structures and Materials, Faculty of Aerospace Engineering, Delft University of Technology, r.debreuker@tudelft.nl, Associate Fellow AIAA} and
	Saullo G.P. Castro \footnote{Associate Professor, Aerospace Structures and Materials, Faculty of Aerospace Engineering, Delft University of Technology, s.g.p.castro@tudelft.nl}
}
\affil{Delft University of Technology, Delft, The Netherlands}

\begin{document}
\thispagestyle{plain}
\maketitle


\begin{abstract}
	Design optimisation potentially leads to lightweight aircraft structures with lower environmental impact. Due to the high number of design variables and constraints, these problems are ordinarily solved using gradient-based optimisation methods, leading to a local solution in the design space while the global space is neglected. Bayesian Optimisation is a promising path towards sample-efficient, global optimisation based on probabilistic surrogate models. While Bayesian optimisation methods have demonstrated their strength for problems with a low number of design variables, the scalability to high-dimensional problems while incorporating large-scale constraints is still lacking. Especially in aeroelastic tailoring where directional stiffness properties are embodied into the structural design of aircraft, to control aeroelastic deformations and to increase the aerodynamic and structural performance, the safe operation of the system needs to be ensured by involving constraints resulting from different analysis disciplines. Hence, a global design space search becomes even more challenging. The present study attempts to tackle the problem by using high-dimensional Bayesian Optimisation in combination with a dimensionality reduction approach to solve the optimisation problem occurring in aeroelastic tailoring, presenting a novel approach for high-dimensional problems with large-scale constraints. Experiments on well-known benchmark cases with black-box constraints show that the proposed approach can incorporate large-scale constraints.
\end{abstract}



\section{Introduction}\label{ch:intro}
\thispagestyle{FirstPage}
Humanity is driving towards a greener future, particularly within sustainable aviation, with an ever-growing demand for more efficient and environmentally friendly aircraft. Improving high-performance aircraft is a crucial step towards a sustainable and widely accepted aviation sector. Achieving this involves optimising structural designs to reduce energy consumption. Therefore, developing new methods and technologies to mitigate environmental impact becomes inevitable.\\
Aeroelastic tailoring is a promising technique for weight reduction in high aspect ratio wings to enhance their performance. Therein, directional stiffness properties are embodied into the structural design of aircraft to control aeroelastic deformations to increase the aerodynamic and structural performance by optimising, for instance, ply thicknesses and angles \parencite{shirk_aeroelastic_1986}. This process involves multiple disciplines, known as multi-disciplinary design optimisation (MDO), ensuring the system's feasibility and safe operation.\\
Evaluating these complex aeroelastic models' computational expense and time-consuming nature demand efficient optimisation algorithms. Gradient-based methods are often utilised in this context to find a local optimum with a reasonable amount of model evaluations. To allow the use of gradient-based optimisation in the aeroelastic tailoring of composite wings, a convenient route makes use of the so-called lamination parameters to describe the laminated composite parts, allowing a condensed description of the membrane, bending, and coupled stiffness terms using continuous variables \parencite{dillinger_stiffness_2013}. With these continuous variables, it is possible to compute gradients in high-dimensional optimisation problems. At the same time, large-scale constraints arise from considering various structural analysis disciplines, such as buckling and other failure criteria.  However, multiple challenges emerge from this gradient-based approach. Firstly, the computation of gradients is not always easy, e.g., if the model's source code is not available, thus relying on approaches like Finite Differences, leading to prohibitive computational costs, which motivates the use of gradient-free methods. Secondly, the response surface for feasible designs in aeroelastic tailoring is known to be multi-modal, trapping gradient-based methods in local optima and neglecting the global design space, potentially hindering the discovery of better designs. Hence, there is a desire to develop methods exploring the global design space, optimising structures to achieve lighter aircraft configurations. \\
The optimisation problem at hand can be formulated as follows:
\begin{equation}\label{eq:generaloptimisationproblem}
	\min_{\textbf{x} \in \mathcal{X}\subset \mathbb{R}^D} f(\textbf{x}) \ \text{s.t.} \forall i \in \{1,...,G\}, c_i(\textbf{x}) \leq 0, 
\end{equation}
where $\mathcal{X}\subset \mathbb{R}^D$ is a $D$-dimensional space of potential designs, $f(\textbf{x}): \textbf{x} \in \mathcal{X} \to \mathbb{R}$ the objective function and $G$ constraints arising from the multi-disciplinary analyses. Overall, this optimisation problem can also be seen as a multi-output problem where the model maps from a vector of design variables to the objective function and $G$ constraints. \\~\\ 
Due to the expensive nature of evaluating the objective function and especially the associated constraints, a sample-efficient algorithm is crucial. Bayesian Optimization (BO) is a powerful method for complex and costly problems, extensively applied across various domains, including aircraft design \textcite{saves_multidisciplinary_2022}. In BO, the expensive-to-evaluate functions that represent the objective and constraints, in many problems seen as a black box, are replaced by a computationally cheap surrogate model by using e.g. Gaussian Processes (GP) \parencite{frazier_tutorial_2018}. While many authors have shown that for lower dimensional problems, BO methods perform well, high-dimensional cases pose significant challenges due to the curse of dimensionality \parencite{eriksson_high-dimensional_2021,priem_optimisation_2020}, resulting from the fact that high dimensional search spaces are difficult to explore exhaustively. However, BO offers a probabilistic approach to efficiently guide through the design space to find promising regions and reduce the computational burden. While these algorithms bring a variety of advantages, their scalability to high-dimensional problems with many constraints, as is often the case in engineering design, remains a great challenge. \\~\\
The present study focuses on employing high-dimensional BO algorithms for aeroelastic tailoring while considering large-scale constraints arising from the multi-disciplinary analyses, as formulated in Equation \ref{eq:generaloptimisationproblem}. First, BO for unconstrained and constrained problems is introduced, and the difficulties in terms of scalability are highlighted. Subsequently, dimensionality reduction in the context of constrained BO is presented before the theory is applied to the aeroelastic tailoring optimisation problem. \\
The novelty of this paper lies in the formulation of a high-dimensional BO method with a dimensionality reduction approach that lowers the computational burden arising from the incorporation of a large number of constraints. Subsequently, the methodology is applied to the $10D$ Ackley function with two black-box constraints as well as to the $7D$ speed reducer problem with 11 black box constraints before some preliminary results are shown for the use in aeroelastic tailoring.\\


\section{High-Dimensional Constrained Bayesian Optimisation}\label{ch:hdbo}
This section briefly introduces Bayesian Optimization (BO) within the context of high dimensionality and constraints. Gaussian Processes (GPs) are introduced as the preferred surrogate modelling technique. Subsequently, GPs are linked to unconstrained BO, which is then expanded to address the constrained scenario, followed by an outline of the challenges encountered in this work.

\subsection{Gaussian Processes}\label{ch:gp}
A Gaussian Process in the context of BO is used to compute a surrogate model which is fast to evaluate and from which the optimum can be obtained cheaply. Recall that $\mathcal{X} \subset \mathbb{R}^D$ is a $D$-dimensional domain and the corresponding minimisation problem is presented in Equation \ref{eq:generaloptimisationproblem}. Starting from a Design of Experiments (DoE), written as $\mathcal{D} = \{ \textbf{x}_i,f(\textbf{x}_i)\}_{i=1,...,N}$, where $\textbf{x}_i \in \mathbb{R}^D$ is the $i$-th of $N$ samples and $f (\textbf{x}_i):\mathcal{X} \to \mathbb{R}$ the objective function, mapping from the design space to a scalar value. Typically, GPs are employed within BO to learn a surrogate model $\hat{f}(\textbf{x}): \mathcal{X} \to \mathbb{R}$ of the objective function $f$ from this given data set $\mathcal{D}$. Therefore, it is assumed that the objective function $f$ follows a GP, also called a multivariate normal distribution $\mathcal{N}$. By defining the mean $m:\mathcal{X} \to \mathbb{R}$ and covariance  $k:\mathcal{X} \times \mathcal{X} \to \mathbb{R}$, the surrogate can be denoted as 
\begin{equation}\label{eq:prior}
	f(\textbf{x}) \sim \hat{f}(\textbf{x}) = \mathcal{GP} \left( m( \textbf{x}) , k(\textbf{x}, \textbf{x}') \right) = \mathcal{N} \left( \mu(\textbf{x}),\sigma( \textbf{x})^2 \right),
\end{equation}
also known as the prior. This encodes the a priori belief that the observations are distributed normally. A common covariance function, sometimes referred to as kernel, is, for instance, the so-called squared exponential kernel $k(\textbf{x},\textbf{x}')$ defined as 
\begin{equation}\label{eq:kernel}
	k(\textbf{x},\textbf{x}') = s^2 \exp{\left( -\frac{1}{2} \sum_{i=1}^{D} \left( \frac{x_i-x_i'}{l_i} \right)^2 \right)},
\end{equation}
encoding the similarity between two chosen points $\textbf{x}$ and $\textbf{x}'$ \parencite{rasmussen_gaussian_2006}. The parameter $l_i$ for $i=1,...,D$ is called the length scale and measures the distance for being correlated along $x_i$. Together with $s^2$, the parameters form a set of so-called hyperparameters $\boldsymbol{\theta} = \{ l_1,...,l_D,s^2\}$ (in total $D+1$ parameters) which need to be determined to train the model with respect to the target function. The kernel matrix is defined as $\textbf{K} = \left[ k(\textbf{x}_i,\textbf{x}_j)  \right]_{i,j=1,...,N} \in \mathbb{R}^{N \times N}$. The kernel needs to be defined such that $\textbf{K}$ is symmetric positive definite to ensure its invertibility. The positive definite symmetry is guaranteed if and only if the used kernel is a positive definite function, as detailed in \textcite{schoenberg_metric_1938}. \\~\\ 
Considering a new query point $\textbf{x}_{*} \in \mathcal{X}$, the stochastic process in Equation \ref{eq:prior} can be used to predict the new query point using Bayes' rule through the posterior distribution, defined as 
\begin{equation}
	f_{*} \sim \mathcal{N} \left( \mu(\textbf{x}_{*}),k(\textbf{x}_{*},\textbf{x}_{*}) \right)
\end{equation}
The posterior mean $\hat{\mu}(\bullet)$ and covariance function $\hat{\sigma}(\bullet)$ are computed with
\begin{align}\label{eq:posterior} 
	\hat{\mu}(\textbf{x}_{*}) &= \textbf{k}(\textbf{x}_{*},\textbf{x}) \textbf{K}(\textbf{x},\textbf{x})^{-1} \textbf{f}, \\
	\hat{\sigma}(\textbf{x}_{*}) &=  k(\textbf{x}_{*},\textbf{x}_{*})-\textbf{k}(\textbf{x}_{*},\textbf{x}) \textbf{K}(\textbf{x},\textbf{x})^{-1} \textbf{k}(\textbf{x},\textbf{x}_{*}),
\end{align}
where $\textbf{x} = [ \textbf{x}_1,\textbf{x}_2,...,\textbf{x}_N]$ is the collection of samples and $\textbf{f} = [f_1,f_2,...,f_N]$ of computed objective values in $\mathcal{D}$. The values of the hyperparameters $\boldsymbol{\theta}$ are determined by maximising the marginal likelihood. More detailed information can be found in \textcite{rasmussen_gaussian_2006}.

\subsection{Unconstrained Bayesian Optimisation}
Up to this point, the GP has been computed based on the initial samples contained in $\mathcal{D}$. BO now aims to iteratively increase the accuracy of the surrogate model, enriching the DoE while exploring the design space. Thus, leveraging the acquired data, the endeavour is to pinpoint regions where optimal values are anticipated. For a thorough derivation, unconstrained BO is considered first. The problem at hand can be written as 
\begin{equation}
	\min_{\textbf{x} \in \mathcal{X}} f(\textbf{x}).
\end{equation}
An acquisition function $\alpha: \mathcal{X} \to \mathbb{R}$ is used to guide the optimisation through the design space while trading off exploration and exploitation based on the posterior mean and variance defined in Equation \ref{eq:posterior}. The former describes the exploration of the whole design space, whereas the latter tries converging to an optimum based on the data observed. However, a multitude of acquisition functions exist. A popular choice is the so-called Expected Improvement (EI) \parencite{mockus_j_application_1978}, denoted as 
\begin{align}\label{eq:EI}
	\alpha_{\mathrm{EI}}(\textbf{x})=
	\left\{\begin{array}{l}
		0 \quad \text{ if } \hat{\sigma}(\textbf{x})=0 \\
		\left(f_{\min }-\hat{\mu}(\textbf{x})\right) \Phi\left(\frac{f_{\min}-\hat{\mu}(\textbf{x})}{\hat{\sigma}(\textbf{x})}\right)+\hat{\sigma}(\textbf{x}) \phi\left(\frac{f_{\min} - \hat{\mu}(\textbf{x})}{\hat{\sigma}(\textbf{x})}\right) \quad \text{ else } 
	\end{array}\right.
\end{align}
where $\Phi(\cdot)$ and $\phi(\cdot)$ are the cummulative distribution and probability density function, whereas $f_{\min}$ represents the observed minimum. By maximising Equation \ref{eq:EI} over the design space $\mathcal{X}$, the new query point $\textbf{x}_{*}$ can be found \parencite{frazier_tutorial_2018}
\begin{equation}\label{eq:newpoint}
	\textbf{x}_{*} \in \argmax_{\textbf{x} \in \mathcal{X}} \alpha_{\mathrm{EI}}(\textbf{x}).
\end{equation}

\subsection{Constrained Bayesian Optimisation}
Most engineering design problems involve constraints, which can be integrated into the previously introduced BO method. There are plenty of algorithms to do so, e.g. \parencite{gardner_bayesian_2014,gelbart_bayesian_2014,hernandez-lobato_general_2016}. Assuming that the output of a model evaluation at design point $\textbf{x}_i$ is not only the objective function $f(\textbf{x}_i)$ but also contains a mapping from the design space to a collection of $G$ constraints $\textbf{c}(\textbf{x}_i): \mathcal{X} \to \mathbb{R}^G$, the DoE for this case can be written as $\mathcal{D} = \{ \textbf{x}_i, f(\textbf{x}_i), \textbf{c}(\textbf{x}_i) \}_{i=1,..,N}$. The new design point found in Equation \ref{eq:newpoint} needs to lie in the feasible space $\mathcal{X}_f$, written as $\textbf{x}_{*} \in \mathcal{X}_f \subset \mathcal{X}$ where $\mathcal{X}_f := \{ \textbf{x} \in \mathcal{X} \text{ s.t. } \hat{c}_{1:G}(\textbf{x}) \leq 0 \}$.
Among others, \textcite{gardner_bayesian_2014} propose to model each constraint $c_j(\textbf{x}), j=1,...,G$ by an independent surrogate model, the same way as it is done for the objective function 
\begin{equation}\label{eq:prior_c}
	c_i(\textbf{x}) \sim \hat{c}_i(\textbf{x}) = \mathcal{GP} \left( m( \textbf{x}) , k(\textbf{x}, \textbf{x}') \right) = \mathcal{N} \left( \mu(\textbf{x}),\sigma( \textbf{x})^2 \right),
\end{equation}
leading to $G+1$ GP models in total, enabling the extension of Equation \ref{eq:EI} for constrained problems, also referred to as Expected Feasible Improvement (EFI), written as
\begin{equation}
	\alpha_{\mathrm{EFI}} = \alpha_{\mathrm{EI}} \prod_{i=1}^{G} \mathrm{Pr} \left( \hat{c}_i(\textbf{x} \leq 0 )\right).
\end{equation}
Accordingly, within the acquisition strategy, the sub-problem
\begin{equation}\label{ch:newpoint}
	\textbf{x}_{*} \in \argmax_{\textbf{x} \in \mathcal{X}_f \subset \mathcal{X}} \alpha_{\mathrm{EFI}}(\textbf{x})
\end{equation}
has to be solved. This subsection solely aims to introduce the crucial aspects of constrained BO briefly and shall stress the fact that each constraint needs to be modelled via a separate GP model. \\
Of course, a multitude of alternatives to incorporate constraints exists. Among these approaches, for instance, is the use of Thompson Sampling (TS) \parencite{hernandez-lobato_parallel_2017} in the constrained setting, as proposed by \textcite{eriksson_scalable_2019} and employed in the course of this work.  

\subsection{High-Dimensional Problems} 
As presented, BO algorithms consist of two components, namely the GPs to model the surrogate based on Bayesian statistics \parencite{rasmussen_gaussian_2006} and the acquisition function to determine where to query the next point to converge towards the minimiser of the objective function. While these algorithms have been proven very efficient for lower-dimensional problems \parencite{binois_survey_2022}, the scaling to higher dimensions implies some difficulties: \linebreak

\begin{itemize} 
	\item The curse of dimensionality, which essentially states that with increasing number of dimensions, the size of the design space increases exponentially, resulting in an intractable search through the whole design space
	\item As the dimensions increase, so does the number of tunable hyperparameters $\boldsymbol{\theta} \in \mathbb{R}^{D+1}$, leading to a computationally costly learning of the GP
	\item Higher-dimensional problems usually require more samples $N$ to construct the surrogate model accurately. Since the covariance matrix is $\textbf{K} \in \mathbb{R}^{N \times N}$, the inversion of this matrix becomes more costly with a complexity of $\mathcal{O}(N^3)$
	\item Not enough information can be collected, leading to the fact that in the $D$-dimensional hyperspace, most of the samples have a big distance to each other, not allowing for an efficient correlation. 
	\item Acquisition optimisation suffers from large uncertainty in a high-dimensional setting and thus requires more surrogate model evaluation \parencite{binois_survey_2022}
\end{itemize} \leavevmode \linebreak

Different approaches have been used to mitigate the problem of high dimensionality when no or only a few constraints are involved. In \textcite{wang_bayesian_2016}, the authors use random projection methods to project the high-dimensional inputs to a lower dimensional subspace, ending up by constructing the GP model directly on the lower dimensional space, drastically reducing the number of hyperparameters. \textcite{raponi_2020,antonov_2022} use (kernel) Principal Component Analysis on the input space to identify a reduced set of dimensions based on the evaluated samples. Afterwards, the surrogate model is trained in this reduced dimensional space. \textcite{eriksson_high-dimensional_2021} use a hierarchical Bayesian model, assuming that some design variables are more important than others. 
Employing a sparse axis-aligned prior on the length scale will remove dimensions unless the accumulated data tells otherwise.  However, a high computational overhead is shown in \parencite{santoni_2023}. Similarly, decomposing methods exist where the original space is decomposed using additive methods \parencite{kandasamy_high_2016,ziomek_are_2023}. Within the Trust-Region Bayesian Optimisation (TuRBO) algorithm, \parencite{eriksson_scalable_2019} split up the design space in multiple trust regions. These trust regions are defined as hyper-rectangles of size $L \in \mathbb{R}$. The size is then determined with the help of the length scale $l_i$, already defined in Equation \ref{eq:kernel}, and a base length scale $L$ as  
\begin{equation}
	L_i = \frac{l_i L}{\left( \prod_{j=1}^{D} l_j \right)^{1/D}}.
\end{equation}
In this approach, an independent GP model is constructed within each trust region and batched Thompson Sampling (TS)  \textcite{thompson_likelihood_1933} is used as the acquisition function. \\~\\
These methods are all considering unconstrained problem, although \textcite{eriksson_high-dimensional_2021} apply the algorithm for the constrained MOPTA08 \parencite{anjos_m_mopta_2008} problem by using penalty methods. Subsequently, in \textcite{eriksson_scalable_2021}, the TuRBO algorithm is extended to take into account multiple constraints. Here, each constraint is modelled by an independent GP process and batched TS is extended for constrained problems. \\~\\
As shown, to scale BO to high-dimensional problems, strong assumptions have to be made to mitigate the aforementioned obstacles. While the mentioned authors show promising results, an approach of taking into account large-scale constraints, as in the aeroelastic tailoring case where $G>10^3$, is still lack. In this work we choose to use the constrained TuRBO algorithm (Scalable Constrained Bayesian Optimisation, SCBO) for high-dimensional BO. In the following, an extension to this method is presented to mitigate the problem of large-scale constraints.\\

\section{Large-Scale Constrained BO via Reduced-dimensional Output Spaces}\label{ch:large-scale-constraints}
Recall the optimisation problem formulated in Equation \ref{eq:generaloptimisationproblem}. By using constrained BO methods, as shown earlier, each of the $G$ constraints needs to be modelled with an independent GP model, denoted as $\hat{c}_i(\textbf{x})$. This work follows the idea of \textcite{higdon_computer_2008} to construct the surrogates on a lower dimensional output subspace.  
This subspace may be found by using dimensionality reduction methods such as Principal Component Analysis (PCA) \parencite{jolliffe_principal_2016} on the training data in $\mathcal{D}$. An extended version of PCA is the kernel PCA (kPCA), presented by \textcite{scholkopf_nonlinear_1998}. \\
While performing the DoE, apart from the samples $\textbf{x}_i$ and the corresponding objective function $f_i$ also the constraint values $\textbf{c}:\mathcal{X} \to \mathbb{R}^G$ are contained in $\mathcal{D}$, enabling the construction of a matrix
\begin{equation}
	\textbf{C}(\textbf{x})= 
	\begin{bmatrix} \textbf{c}(\textbf{x}_1)^T \\ \textbf{c}(\textbf{x}_2)^T \\ \vdots \\ \textbf{c}(\textbf{x}_N)^T \end{bmatrix} =
	\begin{bmatrix} c_1(\textbf{x}_1) & c_2(\textbf{x}_1) & ... & c_G(\textbf{x}_1) \\ 
		c_1(\textbf{x}_2) & c_2(\textbf{x}_2) & ... & c_G(\textbf{x}_2) \\
		\vdots & \vdots & \ddots & \vdots \\
		c_1(\textbf{x}_N) & c_2(\textbf{x}_N) & ... & c_G(\textbf{x}_N) \end{bmatrix} 
	\in \mathbb{R}^{N \times G}
\end{equation}
with $N$ samples and $G$ constraints. The superscript $T$ denotes the transpose. 

\subsection{Principle Component Analysis (PCA)}
Within PCA, a linear combination with maximum variance is sought, such that 
\begin{equation}
	\textbf{C} \textbf{v} = \lambda \textbf{v}
\end{equation}
where $\textbf{v}$ is a vector of constants. These linear combinations are called the principle components of the data contained in $\textbf{C}$ and can be linked with the Singular Value Decomposition (SVD) \parencite{jolliffe_principal_2002}, leading to 
\begin{equation}
	\textbf{C} = \boldsymbol{\Psi} \boldsymbol{\Sigma} \boldsymbol{\Phi}^T.
\end{equation}
The matrix $\boldsymbol{\Psi} = [ \Psi_1,...,\Psi_r ]\in \mathbb{R}^{N \times r}$ has orthonormal columns which are called the left singular eigenvectors, $\boldsymbol{\Sigma} = diag(\sigma_1,...,\sigma_r) \in \mathbb{R}^{r \times r}$ is a diagonal matrix, containing the eigenvalues and $ \boldsymbol{\Phi} = [\phi_1,...,\phi_r] \in \mathbb{R}^{G \times r}$ contains the corresponding right singular eigenvectors. Here, it is assumed that the SVD automatically sorts the eigenvalues and eigenvectors in descending order. By investigating the eigenvalues in $\boldsymbol{\Sigma}$, and choosing the ones with the $g$ highest values, the truncated decomposition is obtained, consisting of the reduced basis containing $g$ orthogonal basis vectors in $\boldsymbol{\Psi}_g \in \mathbb{R}^{G \times g}$ with $g \ll G$.  The new basis vectors can subsequently be used as a projection $\boldsymbol{\Psi}_g: \mathbb{R}^G \to \mathbb{R}^g$ to project the matrix $\textbf{C}$ onto the reduced subspace $\tilde{\textbf{C}} \in \mathbb{R}^{N \times g}$, written as 
\begin{equation}\label{eq:PCAprojection1}
\tilde{\textbf{C}} = \boldsymbol{\Psi}_g^T \textbf{C} \\
\end{equation}
and for each new vector of constraint values $\textbf{c}_* \in \mathbb{R}^{G}$
\begin{equation}\label{eq:PCAprojection2}
\tilde{\textbf{c}}_* = \boldsymbol{\Psi}_g^T \textbf{c}_*
\end{equation}
Summarising, the $G$ constraints $\textbf{c}(\textbf{x})$ can be represented on a reduced subspace through the mapping $\boldsymbol{\Psi}_g$ while the eigenvalues $\sigma_i$ give an indication about the loss of information, potentially drastically lowering the number of constraints that need to be modelled. A graphical interpretation is depicted in Figure \ref{fig:pca_graphical}. For a more thorough derivation of this method, the reader is referred to \textcite{jolliffe_principal_2016}.

\subsection{Kernel Principle Component Analysis (kPCA)}
While PCA can be seen as a linear dimensionality reduction technique, in \textcite{scholkopf_nonlinear_1998} the authors present an extension, called kernel PCA, using a nonlinear projection step to depict nonlinearities in the data. Similarly to the PCA algorithm, the starting point are the (centred) samples $\textbf{c}_i(\textbf{x}_i) \in \mathbb{R}^{G} \ \forall i \in \{ 1,...,N \}$.\\
Let $\mathcal{F}$ be a dot product space (in the following, also called feature space) of arbitrary large dimensionality. A nonlinear map $\boldsymbol{\phi}(\textbf{x})$ is defined as $\boldsymbol{\phi}: \mathbb{R}^G \to \mathcal{F}$.  This map is used to construct a covariance matrix $\mathcal{C}$ defined as 
\begin{equation}
	\mathcal{C} = \frac{1}{N} \sum_{i=1}^{N} \boldsymbol{\phi}(\textbf{c}(\textbf{x}_i)) \boldsymbol{\phi}(\textbf{c}(\textbf{x}_i))^T.
\end{equation}
The corresponding eigenvalues and eigenvectors in $\mathcal{F}$ are computed by solving
\begin{equation}\label{eq:eig-in-feature-space}
	\mathcal{C}\textbf{v} = \lambda \textbf{v}.
\end{equation}
As stated earlier, since the function $\boldsymbol{\phi}$ maps possibly to a very high-dimensional space $\mathcal{F}$, solving the eigenvalue problem therein may be costly. A workaround is used to avoid computations in $\mathcal{F}$. Therefore, similar to the formulation of the GP models in Section \ref{ch:gp}, a kernel $k: \mathbb{R}^G \times \mathbb{R}^G \to \mathbb{R}$ is defined as
\begin{equation}\label{eq:pcakernel}
	k(\textbf{c}(\textbf{x}_i),\textbf{c}(\textbf{x}_j)) = \boldsymbol{\phi}(\textbf{c}(\textbf{x}_i))^T\boldsymbol{\phi}(\textbf{c}(\textbf{x}_j))
\end{equation}
and the corresponding kernel matrix $\textbf{K}_{ij}  \in \mathbb{R}^{N \times N}$ as
\begin{equation}
	\textbf{K}_{ij} := \left( \boldsymbol{\phi}(\textbf{c}(\textbf{x}_j)), \boldsymbol{\phi}(\textbf{c}(\textbf{x}_j)) \right).
\end{equation}
By solving the eigenvalue problem for non-zero eigenvalues
\begin{equation}
	\textbf{K} \boldsymbol{\alpha} = \lambda \boldsymbol{\alpha}
\end{equation}
the eigenvalues $\lambda_1 \leq ... \leq \lambda_N$ and eigenvectors $\boldsymbol{\alpha}^1, ..., \boldsymbol{\alpha}^N$ are obtained. This part can be seen as the linear PCA, as presented before, although in the space $\mathcal{F}$. To map a test point $\textbf{c}_*(\textbf{x})$ from the feature space $\mathcal{F}$ to the $q$-th principle component $\textbf{v}^q$ of Equation \ref{eq:eig-in-feature-space}, the following relationship is evaluated
\begin{equation}\label{eq:KPCAprojection}
	\left( (\textbf{v}^q)^T \boldsymbol{\phi}(\textbf{c}_*(\textbf{x})) \right) = \sum_{i=1}^{N} \boldsymbol{\alpha}_{i}^{q}( \boldsymbol{\phi}(\textbf{c}(\textbf{x}_i)^T  \boldsymbol{\phi}(\textbf{c}_*(\textbf{x}))) \equiv \tilde{\textbf{c}}_*(\textbf{x}_*).
\end{equation}
A graphical interpretation can be found in Figure \ref{fig:kpca_graphical}. The kernel function in Equation \ref{eq:pcakernel} can also be replaced by another a priori chosen kernel function. Examples of kernels and a more detailed derivation of kPCA can be found in the cited literature. 

\subsection{Dimensionality Reduction for Large-Scale Constraints}
When large-scale constraints are involved, the computational time scales drastically since for each constraint one GP model has to be constructed and trained. Thus, describing the constraints on a lower dimensional subspace allows to significantly lower the computational burden. This idea is based on the work of \textcite{higdon_computer_2008}, who project the simulation output onto a lower dimensional subspace where the GP models are constructed. Other works extended this method then by employing, among others, kPCA as well as manifold learning techniques to account for nonlinearities \parencite{xing_reduced_2015, xing_manifold_2016}. However, the aforementioned authors try to approximate PDE model simulations with high-dimensional outputs, whereas, to the best of the authors' knowledge, the combination of dimensionality reduction techniques for use in high-dimensional BO with large-scale constraints is novel. \\~\\
The methods herein presented are capable of extracting the earlier introduced, most important principle components of available data, reducing the required amount of GP models to $g$ instead of $G$, with $\textbf{v}_j$ as the $j$-th orthogonal basis vector. After projecting the data onto the lower dimensional subspace by using either PCA as in equations \ref{eq:PCAprojection1} or \ref{eq:PCAprojection2} respectively, or kPCA in equation \ref{eq:KPCAprojection}, independent GPs are constructed on the reduced-dimensional output space, formulated as  

\begin{equation}
\tilde{\textbf{c}}_i\sim \hat{\tilde{\textbf{c}}}_{i}  = \mathcal{GP} \left( m_i (\textbf{x}) ,  k_i(\textbf{x}, \textbf{x}') \right) \forall i \in \{1,...,g\}.
\end{equation}

\begin{figure}[h]
	\centering
	\begin{subfigure}{.30\textwidth}
		\centering
		\includegraphics[width=\textwidth]{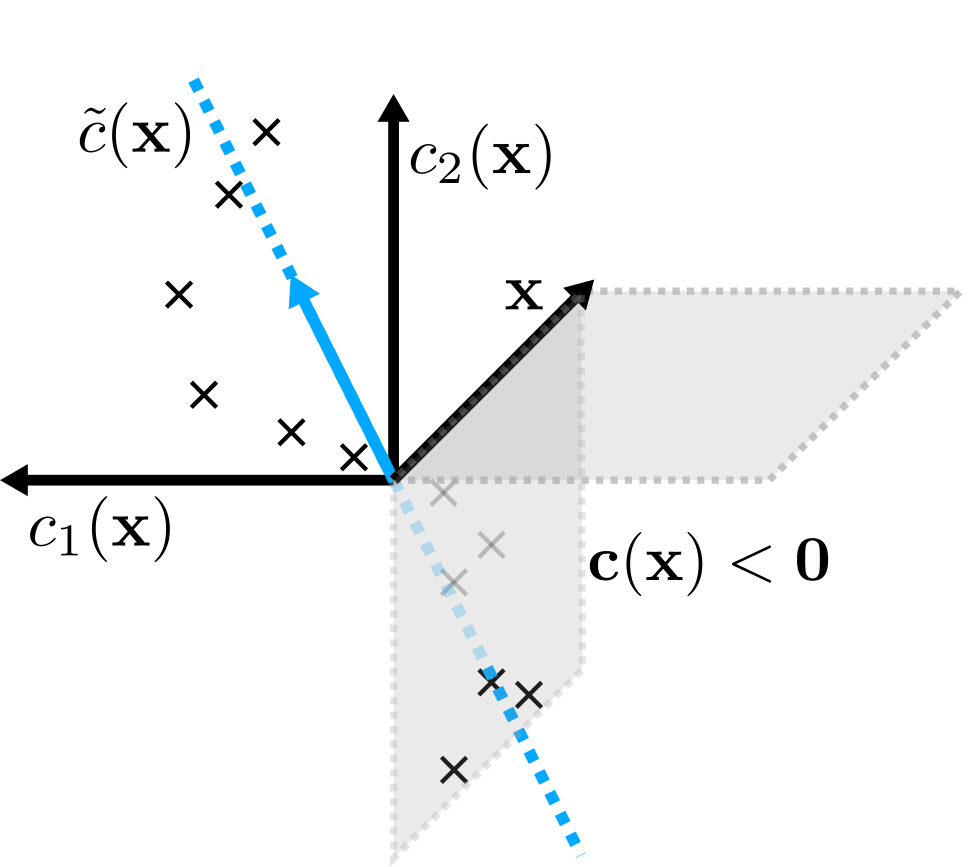}
		\caption{Principal Component Analysis}
        \label{fig:pca_graphical}
	\end{subfigure}
	\hfill
	\begin{subfigure}{.60\textwidth}
		\centering
		\includegraphics[width=\textwidth]{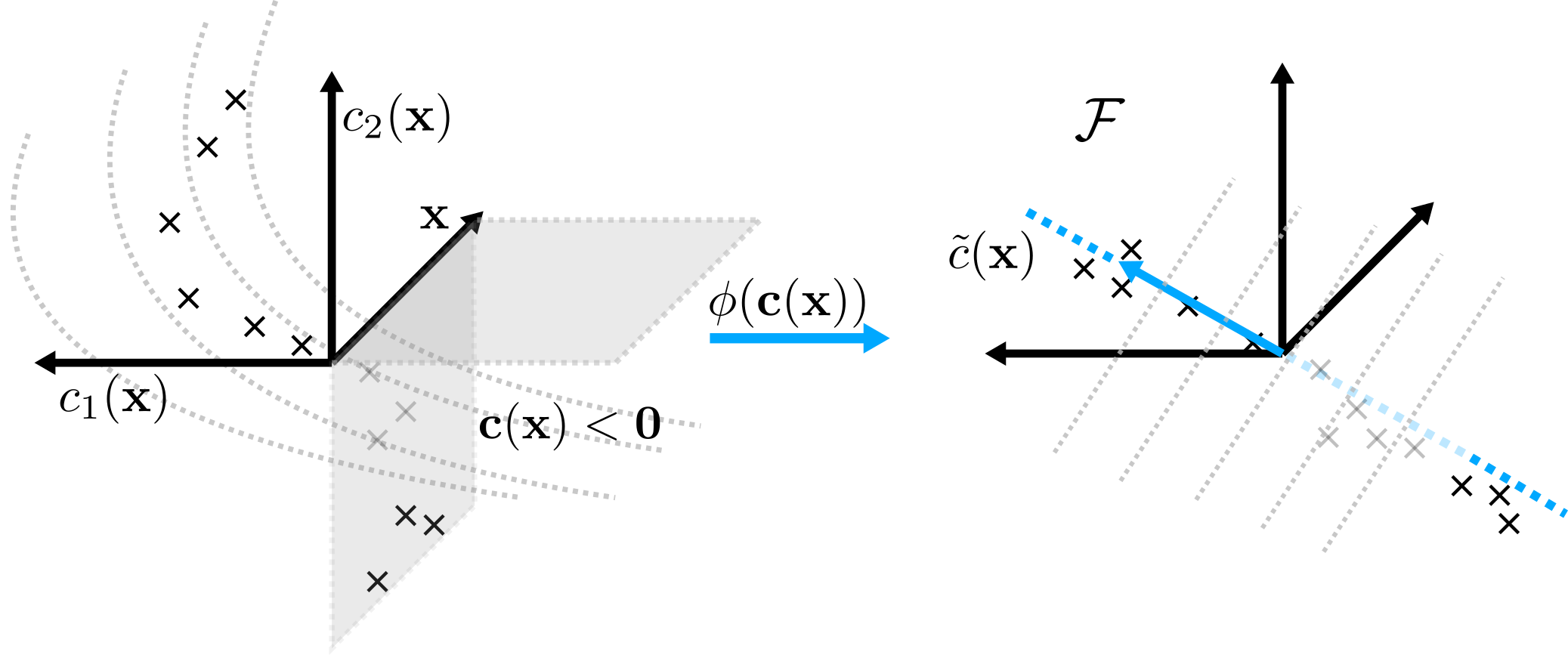}
		\caption{Kernel Principal Component Analysis}
        \label{fig:kpca_graphical}
	\end{subfigure}
	\caption{Graphical interpretation of dimensionality reduction for constraints. On the left, PCA as a linear method is depicted, finding the lower dimensional subspace (blue arrow). On the right, the nonlinear extension, kPCA, is shown, first using a nonlinear kernel to map into the infinite dimensional space $\mathcal{F}$ and subsequently perform the standard PCA. The figure is inspired by \textcite{scholkopf_nonlinear_1998}.}
	\label{fig:pca_kpca}
\end{figure}

A graphical interpretation is depicted in Figure \ref{fig:pca_kpca}.

\section{Aeroelastic Tailoring: A Multi-Disciplinary Design Optimisation Problem}\label{ch:model}
\begin{figure}[h]
	\centering
	\includegraphics[width=10cm]{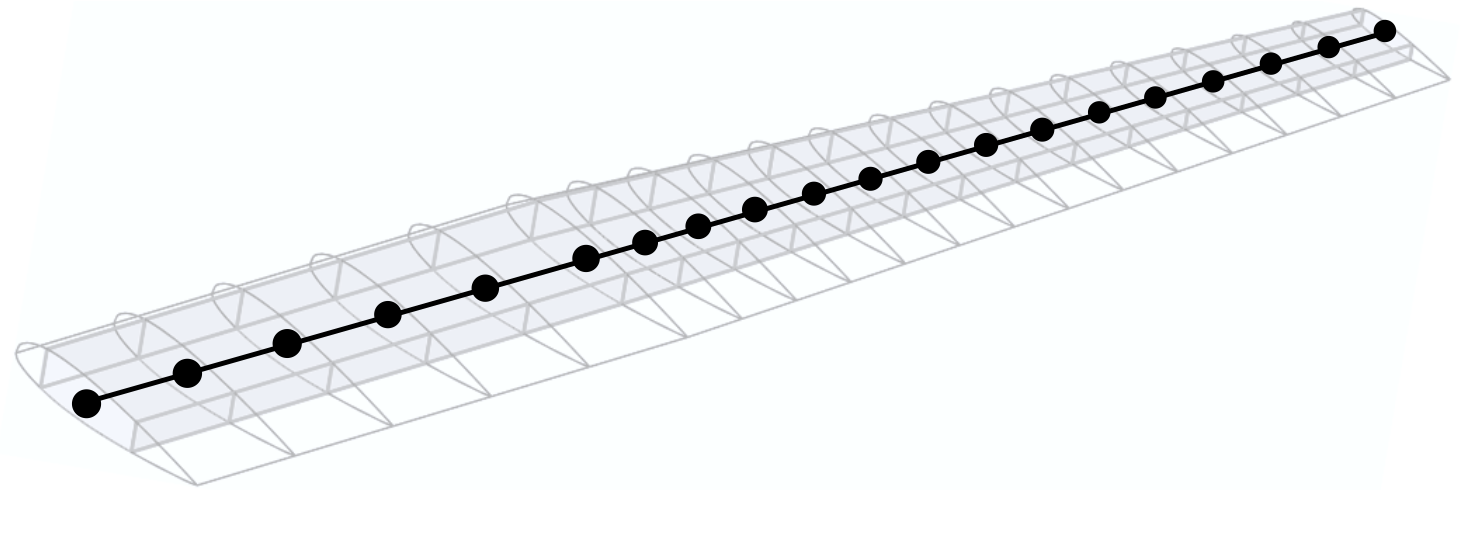}
	\caption{Beam representation of the wing structure}
	\label{fig:beammodel}
\end{figure}

The aeroelastic tailoring model used in this work is based on the preliminary design framework called \textit{Proteus} developed by the Delft University of Technology, Aerospace Structures and Materials, initiated with the work of \textcite{de_breuker_energy-based_2011}. In general, the framework transforms a three-dimensional wingbox made of laminate panels into a three-dimensional beam model, depicted in Figure \ref{fig:beammodel}. A panel in this setting can be, for instance, the upper and lower skin cover or the front and rear spars. The design regions can be discretised by means of various panels along the chord-wise and span-wise directions, ultimately defining the number of design variables.

\begin{figure}[h]
	\centering
	\includegraphics[width=10cm]{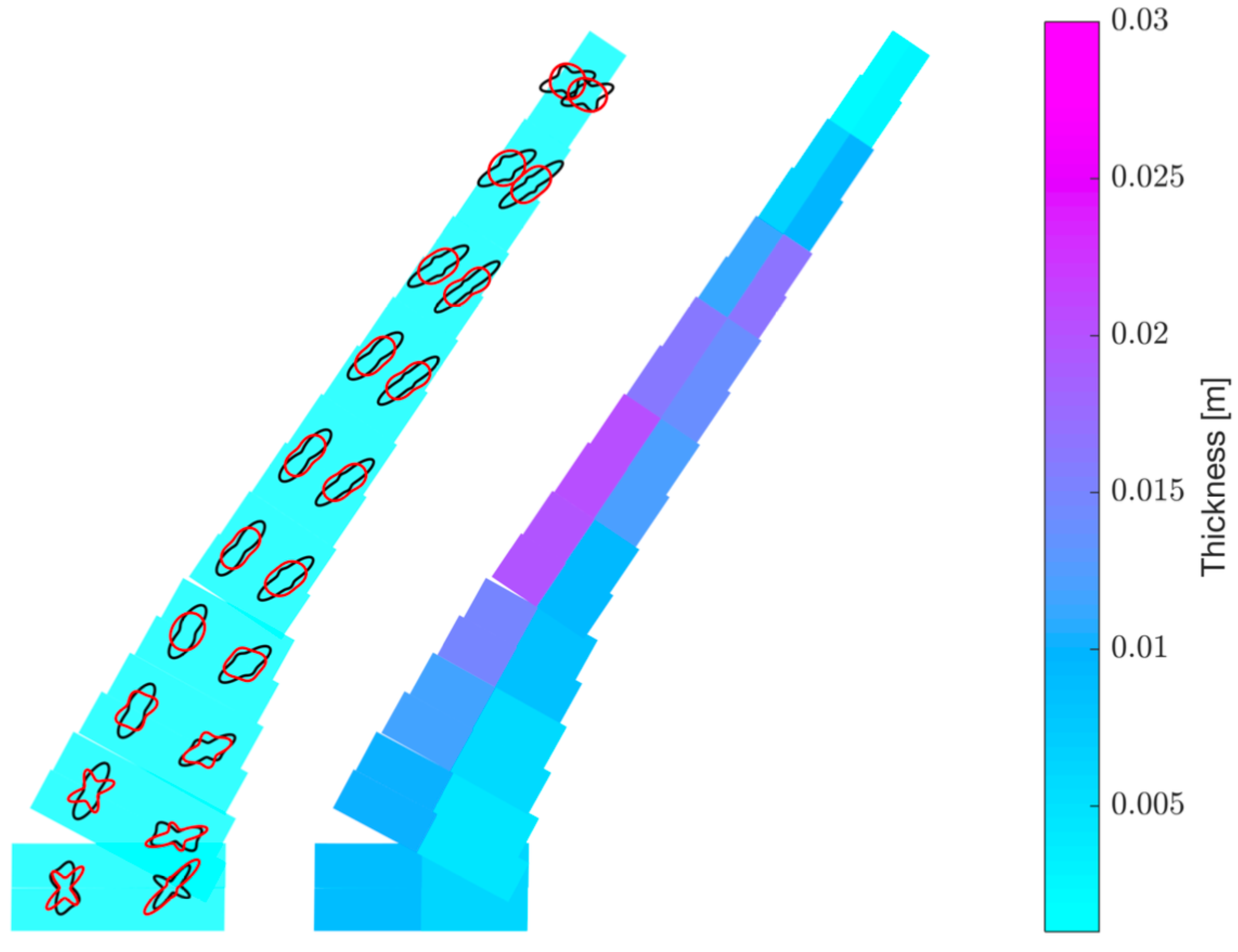}
	\caption{Stiffness and thickness distributions (In-plane stiffness: black, out-of-plane stiffness: red), adopted from  \textcite{rajpal_dynamic_2021}}
	\label{fig:panels}
\end{figure}

The design variables consist of the lamination parameters and the thickness of each panel, respectively, denoted with the superscripts "$lam$" and "$t$". Note that, in the present study the lamination parameters are denoted by $\textbf{x}^{lam}_i$, whereas in the composite community they are often referred to as $\xi_{1,2,3,4}^{A,D}$.
\begin{equation}\label{eq:desvars}
	\textbf{x} = \biggl\{ \textbf{x}^{lam}_1, x^t_1, ...,  \textbf{x}^{lam}_{n_p}, x^t_{n_p} \biggr\}.
\end{equation}
For the sake of illustration, the single panels and the locally optimised thickness and stiffness are depicted in Figure \ref{fig:panels} where a chord-wise panel discretisation of 2 is chosen.
Recall the already introduced optimisation problem from Equation \ref{eq:generaloptimisationproblem}. As mentioned earlier, to ensure the safe operation and feasibility of the system, constraints are computed based on different analysis methods discussed in the following. For the sake of illustration, these constraints are compactly derived here. \\~\\
The lamination parameter feasibility constraints $\textbf{c}_{lpf}$ ensure that the laminates satisfy certain interdependencies and are analytic equations, derived in \parencite{raju_further_2014}, ending up in 6 inequality constraints per panel 
\begin{equation}\label{eq:c-lf}
\textbf{c}_{lpf} = \begin{bmatrix} \textbf{g}_1^T(\textbf{x}) \ \textbf{g}_2^T(\textbf{x}) \ \textbf{g}_3^T(\textbf{x}) \ \textbf{g}_4^T(\textbf{x}) \ \textbf{g}_5^T(\textbf{x}) \ \textbf{g}_6^T(\textbf{x}) \end{bmatrix}^T \leq \textbf{0}.
\end{equation}
Note that each of the six constraints is evaluated for all $n_p$ panels, thus $\textbf{g}_i: \mathcal{X} \to \mathbb{R}^{n_p}$ and $\textbf{c}_{lpf}: \mathcal{X} \to \mathbb{R}^{6n_p}$.
The lamination parameters can be used with the classical laminate theory to construct the following relationship
\begin{equation}
\begin{bmatrix} N \\ M \end{bmatrix} =\begin{bmatrix} \textbf{A}(\textbf{x}) & \textbf{B}(\textbf{x}) \\ \textbf{B}(\textbf{x}) & \textbf{D}(\textbf{x}) \end{bmatrix} \begin{bmatrix} \epsilon^0 \\ \kappa \end{bmatrix}
\end{equation}
This relationship encodes the dependency of the design variables $\textbf{x}$ with the stiffness of the system. A cross-section modeller \parencite{ferede_cross-sectional_2014}, based on a variational approach, is used to obtain the element Timoshenko cross-sectional stiffness matrix $\textbf{C} \in \mathbb{R}^{6 \times 6}$ by relating the strains $\boldsymbol{\epsilon}$ to the applied forces and moment $\boldsymbol{\sigma}$ or $F_i,M_i$ respectively, as in 
\begin{equation}
\boldsymbol{\sigma} = \textbf{C} \boldsymbol{\epsilon} \rightarrow \begin{bmatrix} F_1 & F_2 & F_3 & M_1 & M_2 & M_3 \end{bmatrix}^T =   \textbf{C} \begin{bmatrix} \epsilon_{11} & \epsilon_{12} & \epsilon_{13} & \kappa_{1} & \kappa_{2} & \kappa_{3} \end{bmatrix}^T
\end{equation}
with $\kappa_{1}$ as the twist and $\kappa_{2}, \kappa_{3}$ the bending curvatures. Therewith, the properties of a $2D$ cross section are mapped onto the corresponding beam element node, leading to a $6 \times 6$ element Timoshenko beam stiffness matrix. The corresponding beam strain energy in the continuous form can be derived like 
\begin{align}
\mathcal{U} = \frac{l_0}{2} \int_{0}^1  \boldsymbol{\epsilon}^{T} \textbf{C}  \boldsymbol{\epsilon} d\xi.
\end{align}
By discretisation and introduction of the element degrees of freedom $\textbf{p}$, the linear constitutive stiffness matrix of the beam element, at this point neglecting geometric and material nonlinearities, can be computed by 
\begin{align}
\textbf{K}_{ij} = \frac{\partial^2 \mathcal{U}}{\partial \textbf{p}_i \partial \textbf{p}_j } && \boldsymbol{\epsilon} = \textbf{B} \textbf{p}.
\end{align}
where the matrix $\textbf{B}$ interpolates the nodal quantities to strains $\boldsymbol{\epsilon}$. Please note that in this paragraph $\textbf{K}$ denotes the structural stiffness and not the covariance matrix. Thus, each beam element has $12$ Degrees of Freedom (DoF), $6$ DoF per node. The beam model is depicted in Figure \ref{fig:beammodel}. \\
Assuming a force vector $\textbf{f}$, the static solution $\textbf{p}$ is calculated from
\begin{equation}
\textbf{K}(\textbf{p}) \textbf{p} = \textbf{f}.
\end{equation}
In this framework, geometrical nonlinearities are introduced by using the co-rotational framework of \parencite{battini_co-rotational_2002}, decomposing large displacements/rotations into rigid body displacements and small elastic deformations, ultimately leading to a dependence of the stiffness matrix $\textbf{K}$ on the displacements $\textbf{p}$. After formulating the nonlinear structure, the aerodynamic forces and moments are computed via the unsteady vortex lattice method (UVLM) and mapped onto the structure, resulting in an overall nonlinear aeroelastic system. Due to this nonlinearity, no guarantee exists of finding an equilibrium point right away, motivating the need for an iterative solution to obtain the nonlinear static response. Starting with
\begin{equation}
\textbf{f}_s(\textbf{p}) = \textbf{f}_{ext}(\textbf{p}),
\end{equation} 
where the subscript $s$ denotes the structural force, a predictor-corrector Newton-Raphson solver is used to solve the nonlinear system given by 
\begin{equation}
\left( \frac{\partial \textbf{f}}{\partial \textbf{p}} - \frac{\partial \lambda \textbf{f}_{ext}}{\partial \textbf{p}} \right) \delta \textbf{p} = \textbf{f} - \textbf{f}_{ext} = \textbf{R}.
\end{equation} 
The solution can then be used to compute the corresponding stresses that are used to calculate the Tsai-Wu failure criterion $\textbf{w}(\boldsymbol{\sigma})$, used to assess the strength of the structure. To reduce the number of constraints, only the 8 most critical Tsai-Wu strain factors per panel are considered \parencite{werter_aeroelastic_2017}, leading to
 \begin{equation}\label{eq:c-tw}
\textbf{c}_{tw} = \textbf{w}_{crit}(\boldsymbol{\sigma}) \leq 0. 
\end{equation} 
The buckling analysis assumes that no global buckling can occur due to sufficient strength of stiffeners and ribs. By additionally computing the geometric stiffness matrix $\textbf{K}_g$, the buckling factor $\lambda_b$ can be found by solving the following eigenvalue problem 
 \begin{equation}
\left( \textbf{K} + \lambda_b \textbf{K}_g \right) \textbf{a} = \textbf{0}.
\end{equation} 
To further reduce the number of constraints, only the eight most critical buckling eigenvalues per panel are formulated as a constraint, leading to
 \begin{equation}\label{eq:c-b}
\textbf{c}_{b} = -\boldsymbol{\lambda}_{b,crit} + 1 \leq 0.
\end{equation} 
To compute the aeroelastic stability of the system, the equilibrium between the internal forces and moments $\textbf{f}$ and all the external forces and moments must be regarded. The external forces are split up into applied aerodynamic loads $\textbf{f}_a$ and remaining external forces due to e.g. gravity or thrust $\textbf{f}_e$, given by 
\begin{equation} \label{eq:force-equil}
\textbf{f}_s - \textbf{f}_{ext} = \textbf{f}_s - \textbf{f}_a - \textbf{f}_e = \textbf{0}.
\end{equation} 
By linearising Equation \ref{eq:force-equil}, the corresponding stiffness matrices $\textbf{K}_a$, $\textbf{K}_e$ and $\textbf{K}_s$ can be obtained, and the stability of this static aeroelastic equilibrium is governed by
\begin{equation}\label{eq:static_equil}
\left( \lambda_s \textbf{K}_a + \textbf{K}_e - \textbf{K}_s \right) \Delta \textbf{p} = \textbf{0}.
\end{equation} 
To ensure the static aeroelastic stability of the system, thus preventing divergence, the eigenvalues need to lie in $\lambda_s \geq 1$.  \\
The dynamic aeroelastic analysis is carried out by linearising the system around the static aerodynamic equilibrium and by using the state space formulation for both the aerodynamic and the structural part. It should be mentioned that in the present discussion, many steps are left out for the sake of compactness. More details can be found in \textcite{werter_aeroelastic_2017, de_breuker_energy-based_2011}. As a result, the well-known continuous-time state-space equation is obtained, which can be written as
\begin{equation}
\dot{\textbf{s}} = \textbf{A} \textbf{s} + \textbf{B} \boldsymbol{\alpha}_{air}
\end{equation} 
with $\boldsymbol{\alpha}_{air}$ being the perturbation angle of attack of the induced free stream flow. For dynamic aeroelastic stability, used to prevent flutter, the eigenvalue problem on the state matrix $\textbf{A}$ has to be solved once again, written as 
\begin{equation}
\left( \textbf{A} - \textbf{I} \lambda_f \right) \boldsymbol{\varphi} = \textbf{0}.
\end{equation} 
Anew, only the ten most critical eigenvalues $\lambda_{f,crit}$ are considered \parencite{werter_aeroelastic_2017}, leading to
\begin{equation}\label{eq:c-ds}
\textbf{c}_{ds} = \Re(\lambda_{f,crit}) \leq 0.
\end{equation} 
Furthermore, two more types of constraints are formulated. The aileron effectiveness is constrained as follows
\begin{align}\label{eq:c-ae}
	c_{ae} = \eta_{eff} - \eta_{eff,min} \leq 0,
\end{align}
to ensure safe manoeuvrability of the aircraft as well as the angle of attack $\alpha$ is constrained by using an upper and lower bound, written as 
\begin{equation}
    \begin{aligned}\label{eq:c-aoa}
    \textbf{c}_{AoA,lb} &= -\alpha - \alpha_{lb} \leq 0, \\
    \textbf{c}_{AoA,ub} &= \alpha - \alpha_{ub} \leq 0
    \end{aligned} 
\end{equation}
adding two more constraints per aerodynamic cross-section. \\
Finally, the constraints can be concatenated to form together with the objective function $f(\textbf{x})$ the outputs of the model, as introduced in Section \ref{ch:large-scale-constraints}, written as $\textbf{c}(\textbf{x}) = \{ \textbf{c}_{lpf}, \textbf{c}_{tw}, \textbf{c}_{b},\textbf{c}_{ds}, c_{ae}, \textbf{c}_{AoA} \}^T$. As depicted in Table \ref{table:constraints}, all categories of constraints besides the lamination parameter feasibility need to be taken into account per load case; thus, with an increasing number of loading conditions, the number of constraints quickly increases to a magnitude of $10^3-10^5$.\\
This section aims to expose the origin of the constraints. Nevertheless, it is not always easy or even possible to obtain gradients of those, which is why gradient-free methods such as the Bayesian optimisation, herein proposed, can be very useful.  

\begin{table}[!h]
	\caption{Aeroelastic Tailoring constrained optimisation problem}
	\centering
	\begin{tabular}{c c c c c c}
		\hline\hline
		\textbf{Type} & \textbf{Parameter} & \textbf{Symbol}& \textbf{Equation} &\textbf{/Loadcase} & \textbf{}\\
		\hline
		Objective & Minimise Wing Mass & $f$ & & \\
		\hline
		\multirow{2}{*}{Design Variables ($D$)} & Lamination Parameter & $\textbf{x}_i^{lam}$ & \multirow{2}{*}{(\ref{eq:desvars})} &\multirow{2}{*}{}\\
		& Laminate Thickness & $\textbf{x}_i^{t}$ & \\
		\hline
		\multirow{6}{*}{Constraints ($G$)} & Laminate Feasibility & $\textbf{c}_{lpf}$ & (\ref{eq:c-lf}) & No & Analytic \\
		& Static Strength & $\textbf{c}_{tw}$ & (\ref{eq:c-tw}) & Yes & Analysis \\
		&  Buckling & $\textbf{c}_{b}$ & (\ref{eq:c-b}) & Yes & Analysis\\
		& Aeroelastic Stability & $\textbf{c}_{ds}$ & (\ref{eq:c-ds}) & Yes & Analysis \\
		& Aileron Effectiveness & $c_{ae}$ & (\ref{eq:c-ae}) & Yes & Analysis \\
		& Local Angle of Attack & $\textbf{c}_{AoA}$ & (\ref{eq:c-aoa}) & Yes & Analysis \\
		\hline\hline
	\end{tabular}
	\label{table:constraints}
\end{table}

\section{Application}\label{ch:application}
In this section the presented methodology is applied to two well-known benchmark cases before preliminary results for the aeroealastic tailoring optimisation problem are shown. For the sake of comparison, we follow the same approach as \textcite{eriksson_scalable_2021} and \textcite{hernandez-lobato_general_2016}. Any feasible solution is preferred over an infeasible one. That is why the maximum value of all found feasible solutions is taken as the default value for all infeasible solutions and noted as a dotted red line. Moreover, all computations are performed on an Apple M1 Pro chip while using the frameworks BoTorch \parencite{balandat2020botorch} and GPyTorch \parencite{gardner2018gpytorch}.

\subsection{Academic Example: 10D Ackley Function with 2 Black-Box Constraints}\label{ch:ackley}
The in Section \ref{ch:large-scale-constraints} presented methodology is employed on the well-known Ackley function. This problem has a dimensionality of $D=10$. Additionally, 2 black-box constraints are considered. The optimisation is performed within the domain $[-5,10]^{10}$, and can be written as 

\begin{align}
    f(\textbf{x}) &= -20 \exp \left( -0.2 \sqrt{\frac{1}{d} \sum_{d}^{i=1} x_i^2} \right) - \exp \left( -0.2 \frac{1}{d} \sum_{d}^{i=1} \cos(2 \pi x_i) \right) \\
    c_1(\textbf{x}) &= \sum_{i=1}^{10} x_i \leq 0 \\
    c_2(\textbf{x}) &= ||\textbf{x}||_2 - 5 \leq 0 
\end{align}

As mentioned earlier, the constrained TuRBO algorithm, SCBO, is employed. The same hyperparameters are used as presented in \textcite{eriksson_scalable_2021}, a batch size $q=4$ and $N=10$ initial samples in $\mathcal{D}$. The two constraints are projected onto a lower dimensional subspace (G=2, g=1) using PCA/kPCA (in the following called as SCBO-PCA/SCBO-kPCA). Within SCBO-kPCA the exponential kernel is chosen. In addition, as proposed by \textcite{eriksson_scalable_2021}, a bilog transformation is employed on the constraints to emphasise the region around zero, which is crucial for whether a design is feasible or not. 

\begin{figure}[h]
	\centering
	\includegraphics[width=10cm]{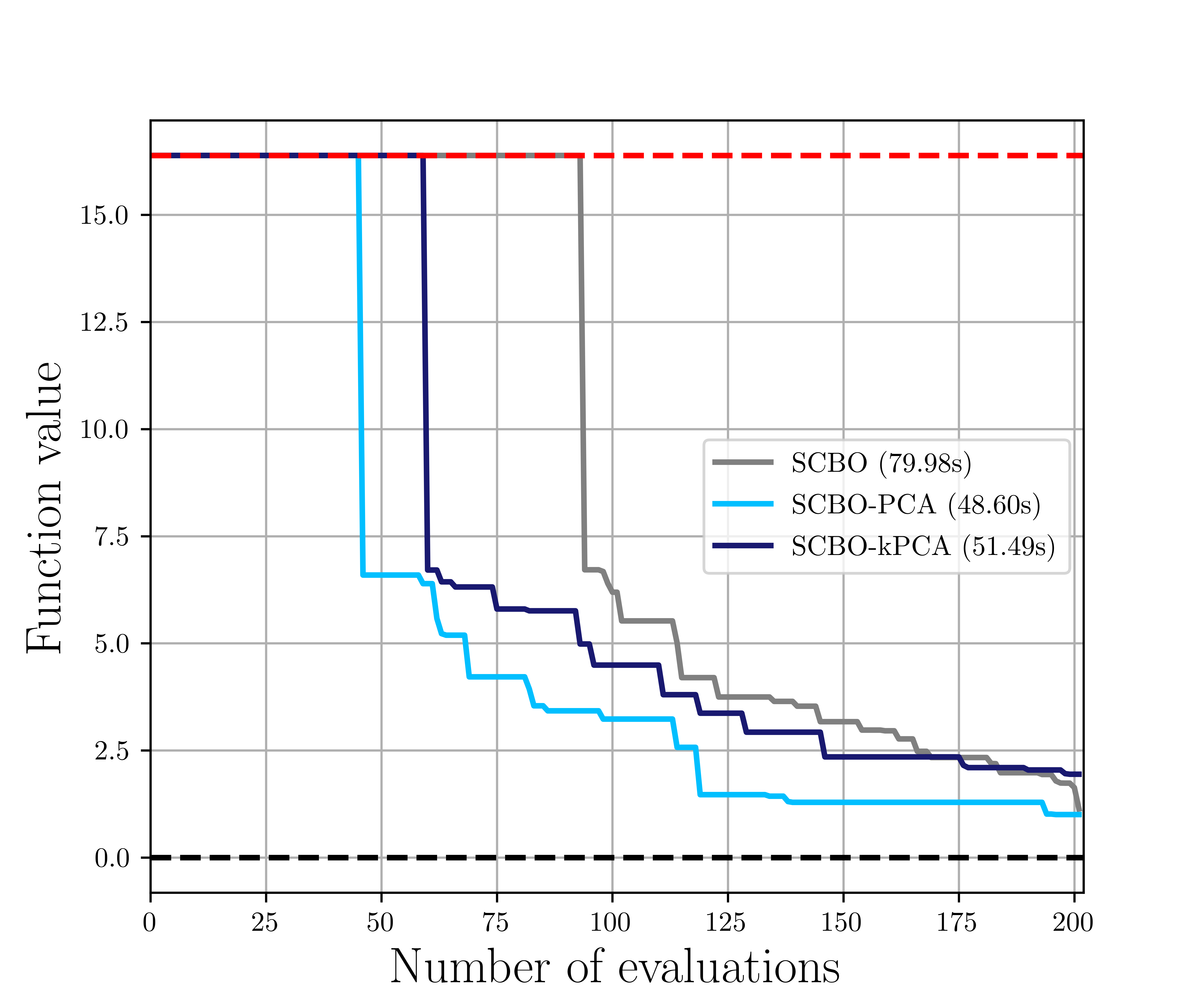}
	\caption{A comparison of the optimisation of the $10D$ Ackley function for SCBO and SCBO combined with PCA and kPCA. Only the objective values for feasible designs are plotted. All methods find a feasible optimum.}
	\label{fig:ackley}
\end{figure}

As it can be seen in Figure \ref{fig:ackley}, even though SCBO-PCA and SCBO-kPCA construct the GP solely on the lower dimensional subspace, the methods still perform as well as or even outperform the standard SCBO while saving approximately 40$\%$ computational time. The performance of SCBO compared to other state-of-the-art optimisation algorithms can be found in \textcite{eriksson_scalable_2021}.

\subsection{Academic Example: 7D Speed Reducer Problem with 11 Black-Box Constraints}\label{ch:speedreducer}
\begin{figure}[h]
	\centering
	\begin{subfigure}{.45\textwidth}
	\centering
	\includegraphics[width=\textwidth]{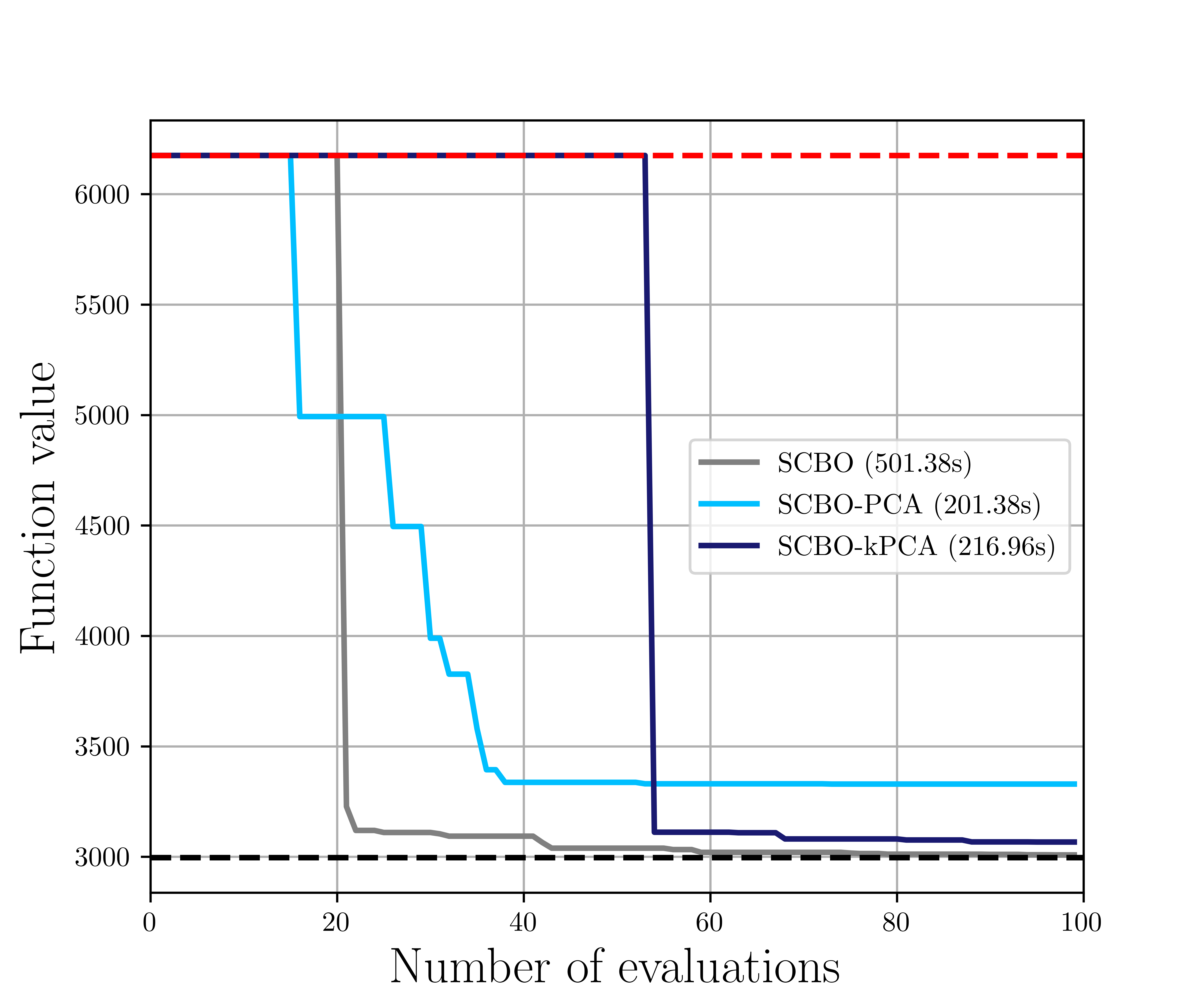}
	\caption{}
	\label{fig:speedreducer_comp}
	\end{subfigure}
	\begin{subfigure}{.45\textwidth}
		\centering
		\includegraphics[width=\textwidth]{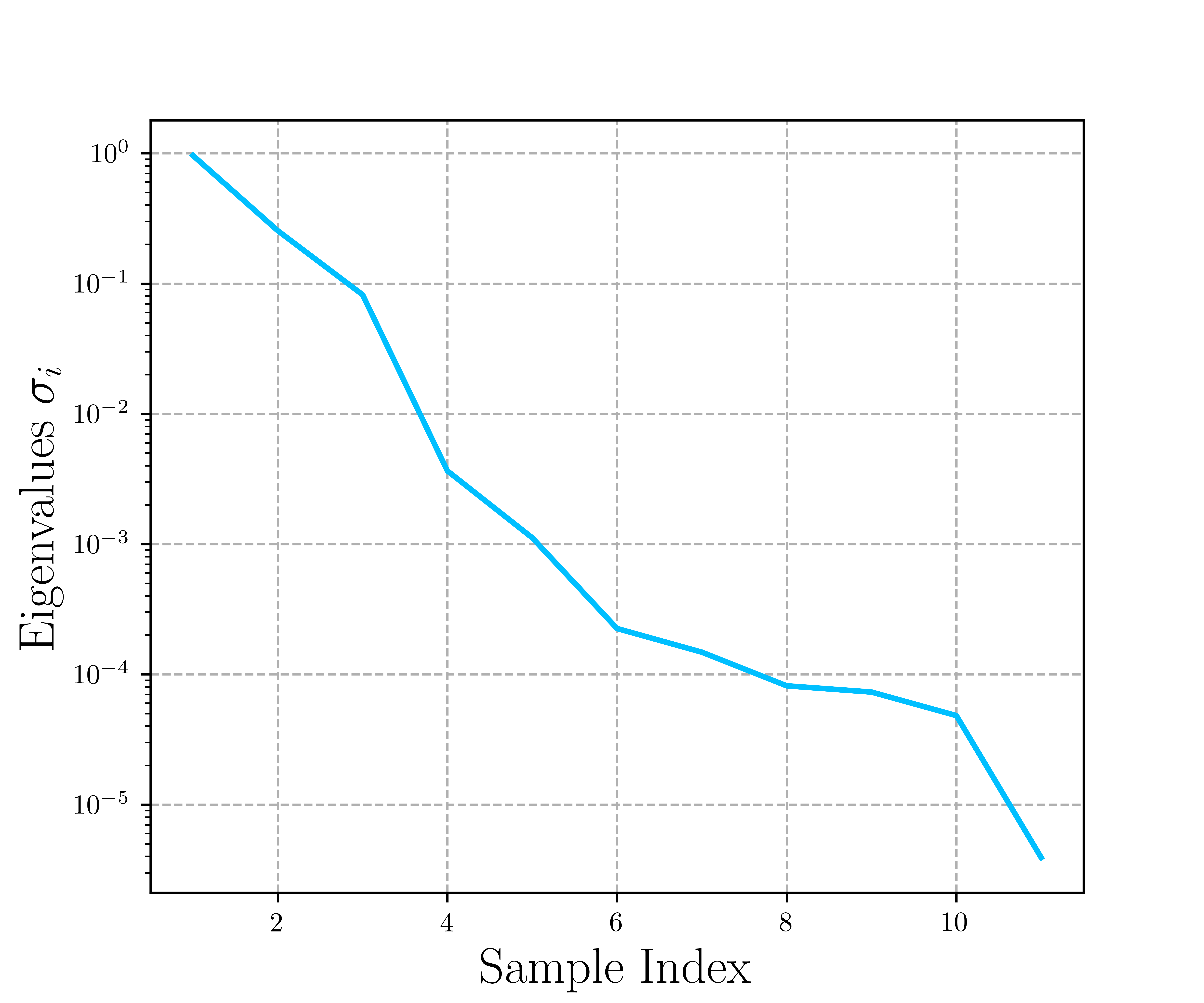}
		\caption{}
        \label{fig:speedreducer_evs}
	\end{subfigure}
	\caption{$7D$ Speed reducer problem with $11$ black-box constraints from \textcite{lemonge_2010}. In (a), SCBO, SCBO-PCA and SCBO-kPCA are compared. In (b), the eigenvalues of the matrix $\textbf{C}$ are plotted}.
    \label{fig:speedreducer}
\end{figure}

Next, the methodology is applied to the $7D$ speed reducer problem from \textcite{lemonge_2010}, including $11$ black-box constraints. The results can be found in Figure \ref{fig:speedreducer}, whereas Figure \ref{fig:speedreducer_comp} shows the decay of the feasible objective values. Figure \ref{fig:speedreducer_evs}, by contrast, depicts the eigenvalues of the constraint matrix $\textbf{C} \subset \mathcal{D}$. In this example, where $G=11$, $g=4$ principal components are chosen. Again, these are the same hyperparameters as presented in Subsection \ref{ch:ackley}, meaning a batch size $q=4$ and $N=10$ initial samples. 

\begin{figure}[h]
	\centering
	\includegraphics[width=10cm]{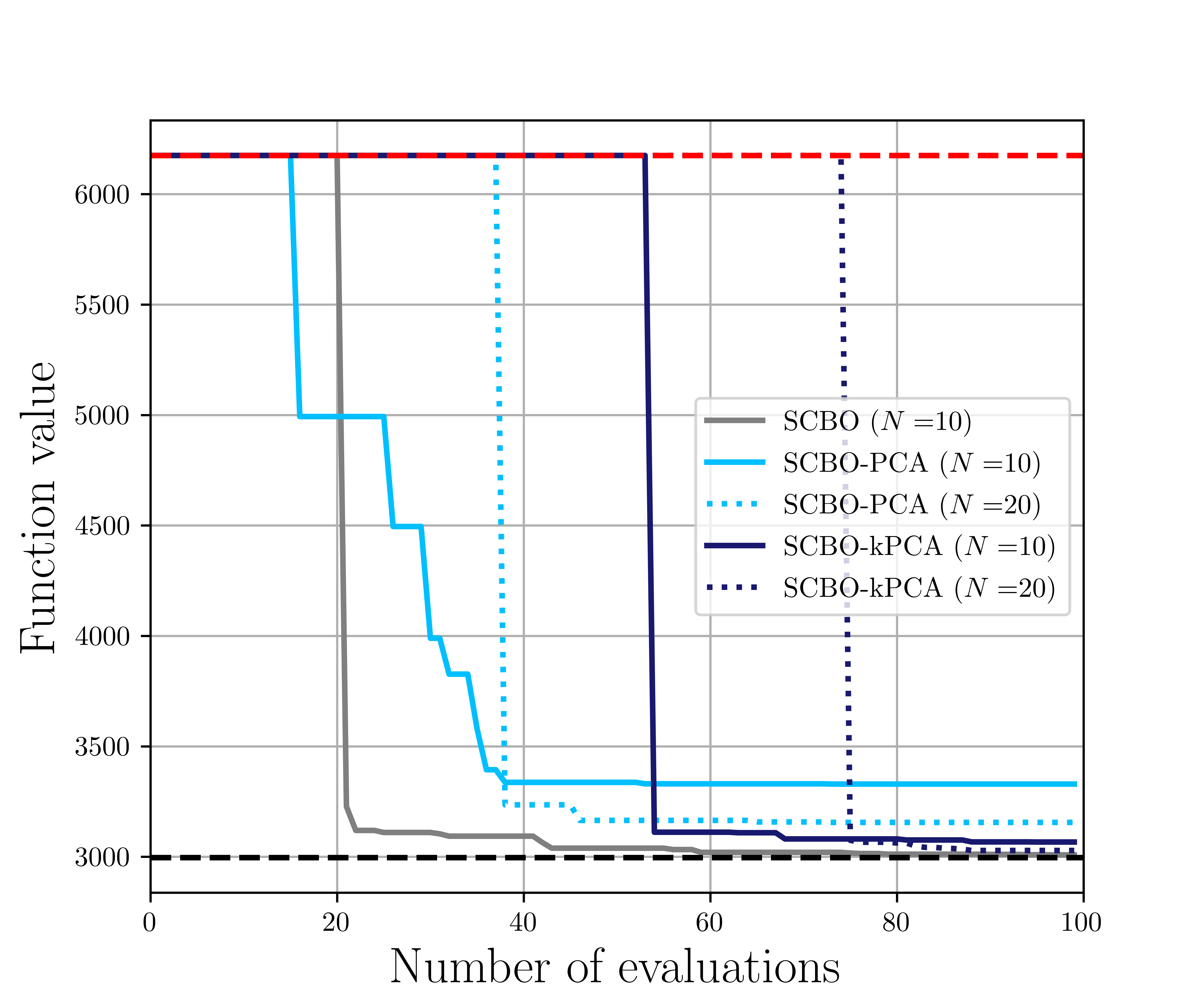}
	\caption{Investigation of number of samples $N$ in $\textbf{C}$ for the $7D$ speed reducer problem with $11$ black-box constraints.}
	\label{fig:speedreducer_doe_inv}
\end{figure}

All methods find a feasible design, whereas SCBO-kPCA performs better than SCBO-PCA. Nevertheless, both methods are significantly faster than the standard SCBO. The better performance of SCBO-kPCA might stem from the ability to capture a nonlinear lower subspace and hence offer a better approximation.\\
However, the lower dimensional subspace is constructed based on the constraint values in $\mathcal{D}$. Assuming that the global optimum lies on the boundary of the feasible space $\mathcal{X}_f$, the success of the method highly depends on how accurately the lower dimensional subspace captures the original space. Hypothesising that the data in $\mathcal{D}$ with $N=10$ was not sufficient, in Figure \ref{fig:speedreducer_doe_inv} the number of initial samples is doubled. 
It can be seen that by increasing the DoE, even better objectives can be found, presuming that due to the additional data, a more accurate subspace and, thus, an optimum closer to the optimum of SCBO can be found. This leads to the conclusion that a sensitivity analysis of the number $N$ might be very important for the method's success. Furthermore, SCBO-PCA and SCBO-kPCA needed more evaluations to find the first feasible design.

\subsection{Aeroelastic Tailoring: A Multi-Disciplinary Design Optimisation Problem}
This work aims to adapt the proposed BO method for the use in aeroelastic tailoring, posing a high-dimensional problem with large-scale constraints, as explained in Section \ref{ch:model}. Assuming that $10^{3} < G < 10^{5}$ high-dimensional GPs are not feasible from a computational standpoint, the presented methodology shall speed up the process by modelling the constraints on a lower-dimensional subspace. The number of design regions has been decreased, ending up at $D=108$, as well as limiting the number of loadcases to one or two, respectively. \\
By using the approach presented in Section \ref{ch:large-scale-constraints}, this work aims to numerically reduce the number of constraints and construct the surrogate models via GPs directly on the lower-dimensional subspace, as demonstrated in Subsections \ref{ch:ackley} and \ref{ch:speedreducer}. In Table \ref{table:constraints}, the sources of constraints are shown. The multitude of outputs arises from the inclusion of multiple loadcases. The premise of this approach lies in the consistency of the physics governing the constraints across loadcases, where eventually only the load changes. This stresses the potential for compressing this information due to the unchanged underlying physics for varying loadcases. 
\begin{figure}[h]
	\centering
	\begin{subfigure}{.45\textwidth}
		\centering
		\includegraphics[width=\textwidth]{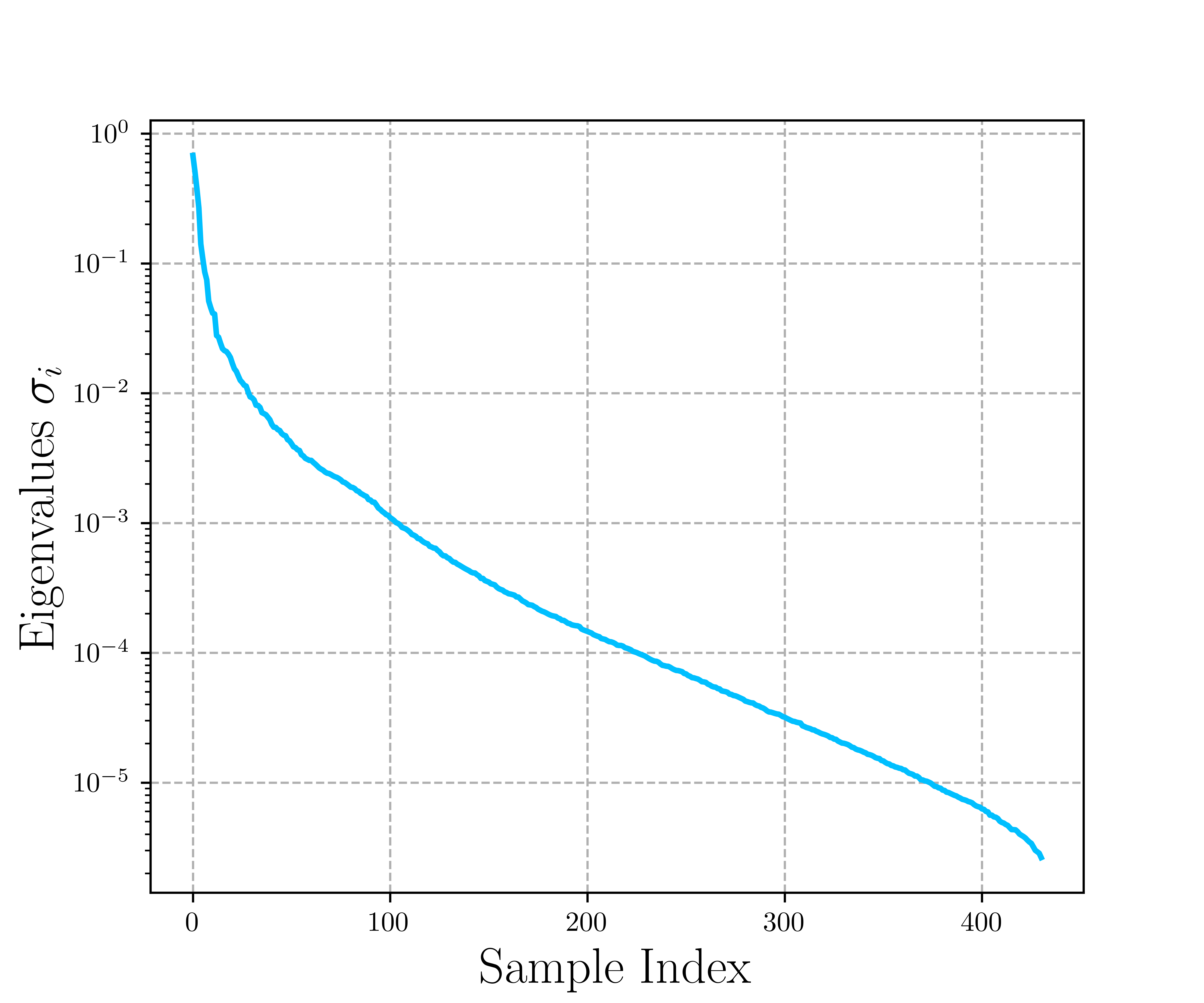}
		\caption{}
        \label{fig:pca}
	\end{subfigure}
	\begin{subfigure}{.45\textwidth}
		\centering
		\includegraphics[width=\textwidth]{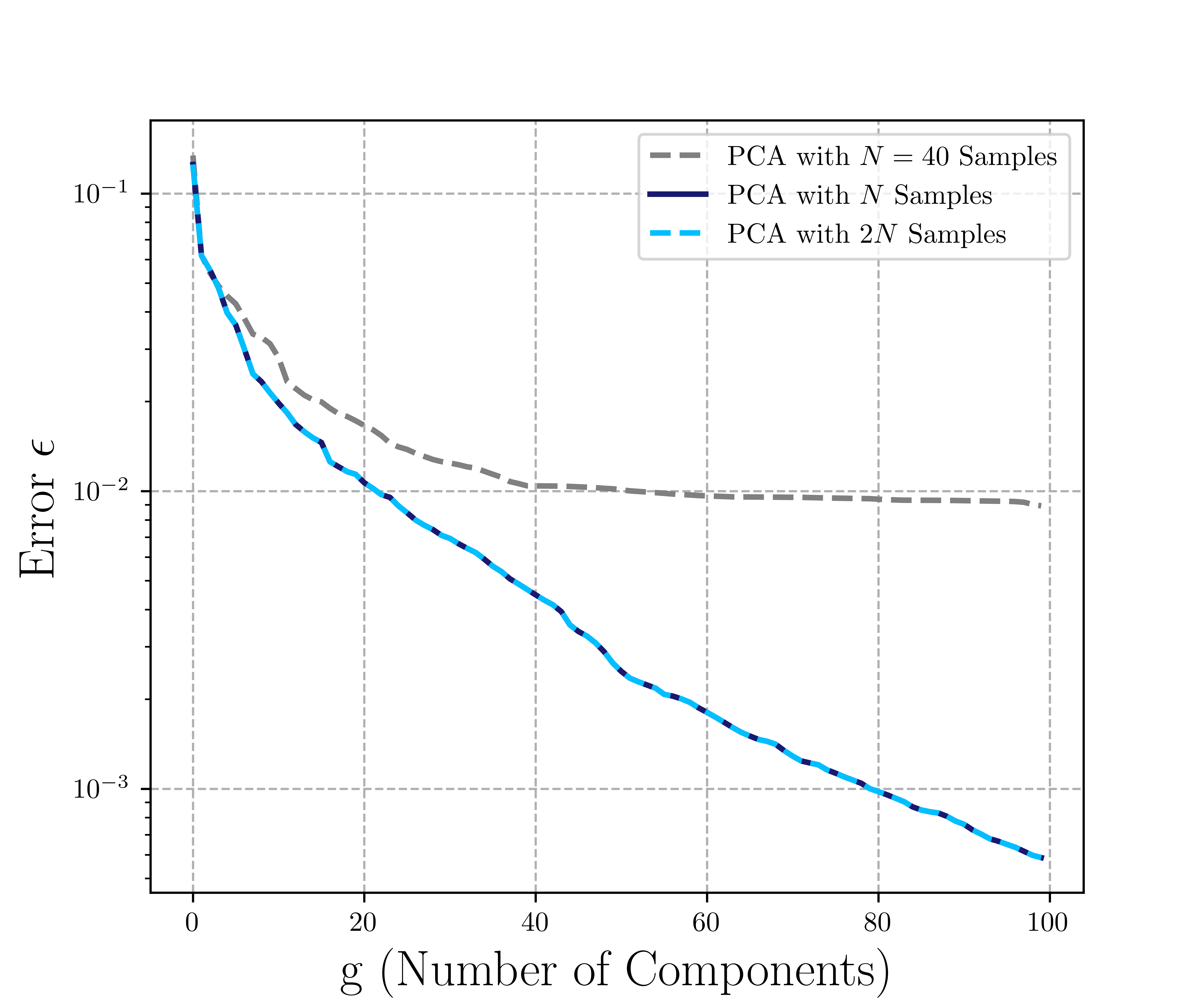}
		\caption{}
        \label{fig:pca_error}
	\end{subfigure}
	\caption{Investigating the constraints in $\mathcal{D}$. (a) shows the decay of the eigenvalues for $N=416$. (b) computes the error depending on how many principal components are taken into account (Equation \ref{eq:error})}.
\end{figure}
The aforementioned aeroelastic tailoring model is used to compute the DoE $\mathcal{D}$ with $N=416$ samples. Sampling was performed via Latin Hypercube Sampling (LHS). Subsequently, PCA is applied on the matrix $\textbf{C}$ to investigate its eigenvalues.  Figure \ref{fig:pca} shows the decay of these computed eigenvalues. If the same error is used as in Subsection \ref{ch:speedreducer}, eigenvalues up to approx $\sigma_i \approx 10^{-2}$, thus $g \approx 29$ principal components might be enough to construct a lower dimensional subspace of sufficient accuracy. In addition, the projection error can be computed. Therefore, some unseen data $\textbf{C}_{*}$, meaning data that has not been used to compute the principal components, is mapped onto the lower dimensional subspace $\tilde{\textbf{C}}_* = \boldsymbol{\Psi}_g^T \textbf{C}_{*}$. Since PCA is a linear mapping, the inverse mapping can be simply computed by  $\hat{\textbf{C}}_{*} = \tilde{\textbf{C}}_* \boldsymbol{\Psi}$. The approximation error can then be computed by
\begin{equation}\label{eq:error}
	\epsilon = \frac{ \parallel \textbf{C}_* - \hat{\textbf{C}}_* \parallel_{F}^{2} }{ \parallel \textbf{C}_* \parallel_{F}^{2}}.
\end{equation}
In Figure \ref{fig:pca_error}, the trend reveals that including more components leads to a reduced error, even for unseen data. Furthermore, to investigate how the construction of the lower-dimensional subspace behaves with sample size variation, the error $\epsilon$ is shown for $N=40$, $N=416$ and $2N$ samples. It can be seen that the error is approximately the same for the latter two cases. As anticipated, an insufficient initial sample size $N$ results in limited information availability during the subspace construction, consequently leading to a larger error. Moreover, the conclusion drawn is that even with $N=416$ samples, sufficient data is available to attain a reasonable subspace. Further, increasing the number of samples in the DoE does not contribute to higher accuracy. \\
As previously noted, the high number of constraints stems from the incorporation of multiple loadcases. Consequently, it becomes intriguing to explore how the eigenvalues vary when the number of loadcases is altered.
\begin{figure}[h]
	\centering
	\includegraphics[width=10cm]{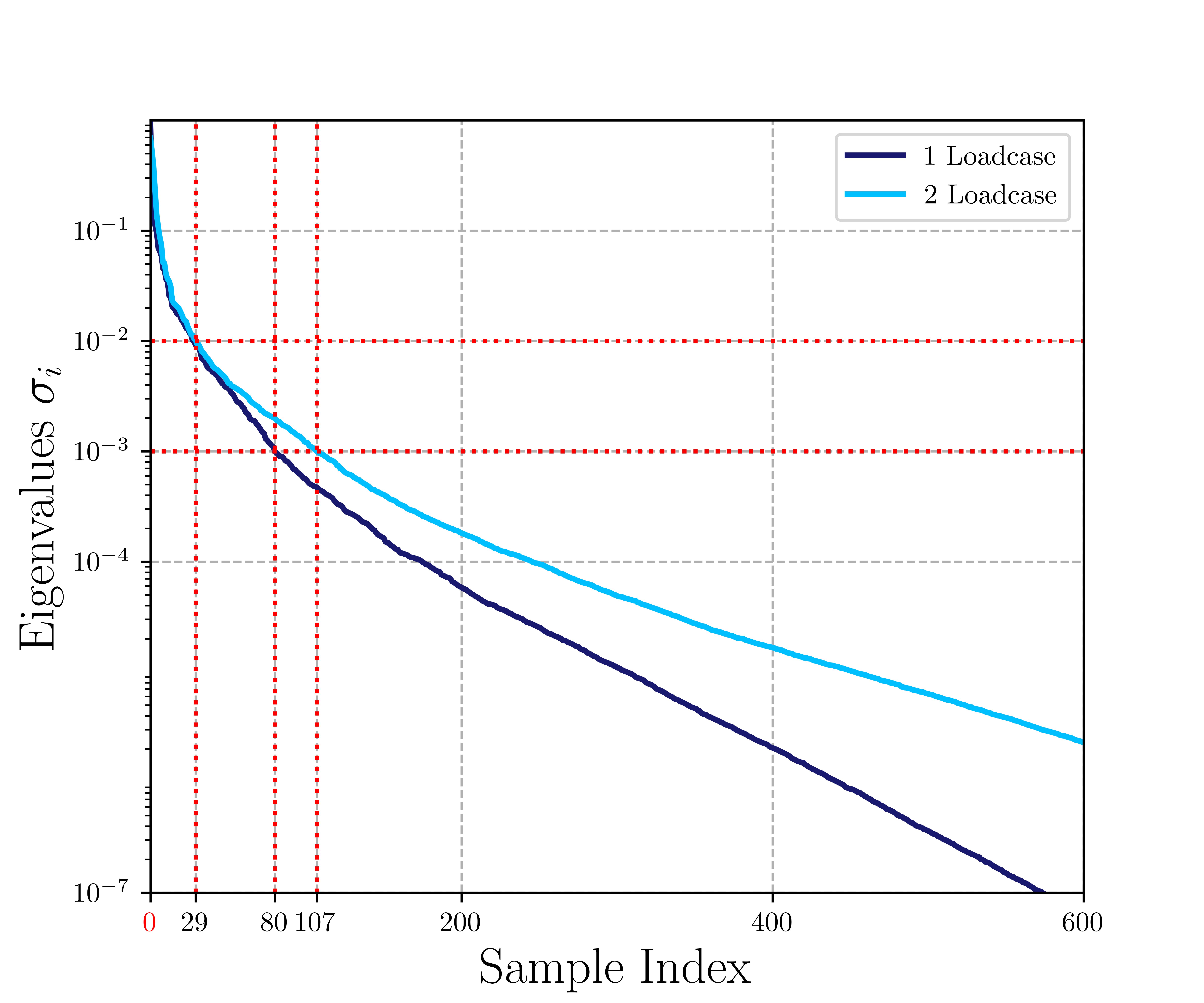}
	\caption{Performing PCA on the matrix $\textbf{C}$ for one and two loadcases.}
    \label{fig:pca_2lc}
\end{figure}
Recall that the eigenvalues denote the importance of their corresponding eigenvector, which serves as a measure of where to truncate the projection matrix. In Figure \ref{fig:pca_2lc}, it can be observed that, even though the number of constraints in the original space has doubled, from $G=893$ to $G=1786$, if the eigenvalues $\sigma_i > 10^{-2}$ are used, no more principal components have to be taken into account. For $\sigma_i > 10^{-3}$, however, only $27$ more components are needed to maintain the same error. Beyond that, the threshold of the eigenvalues is commonly set based on experience, thus can be seen as a hyper-parameter of the method.\\ 
These promising preliminary results motivate the use of the introduced SCBO algorithm \parencite{eriksson_scalable_2021} in combination with a reduced-basis approach to lower the number of constraints in a Bayesian Optimisation to allow a global search of the design space in this high-dimensional problem with large-scale constraints. 

\section{Conclusion and Future Work}
Aeroelastic tailoring can be seen as a high-dimensional multi-disciplinary design optimisation problem with large-scale constraints. Since the global design space search in this use case is not a trivial task, this work uses BO to do so. However, the application of constrained BO is not straightforward due to the poor scalability in terms of the number of constraints. This work introduces a novel approach where a large number of constraints is mapped onto a lower dimensional subspace where the surrogate models are constructed. \\~\\
The herein-presented numerical findings clearly indicate the applicability of this approach. As it can be seen in Section \ref{ch:application}, SCBO with kPCA performs similarly to SCBO while being computationally more efficient. It should be noted that this computational saving can become even more significant when working in a high-dimensional setting where the training of each GP becomes crucial. Thus, by drastically reducing the number of needed GPs, major computational savings can be obtained. \\~\\
Furthermore, this work entails preliminary investigations for the use of this methodology in aeroelastic tailoring, likewise showing promising results. Until now, PCA has been used solely to perform the presented investigations. However, as shown, kPCA can be seen as a nonlinear extension of PCA, which is why even better performance within the optimisation of the aeroelastic tailoring problem is expected due to the nonlinear nature of the constraints. Thus, follow-up studies will investigate these aspects and will aim to incorporate thousands of constraints into the optimisation process. \\~\\
Even though this work has been performed within the realm of aeroelastic tailoring, it is important to stress the generality of the herein-proposed method. As indicated by the numerical investigations, this approach can easily be applied to all sorts of problems where large-scale constraints are involved. \\~\\
To critically reflect the methodology adopted in this work, the following statements can be made. Some authors proposed so-called Multi-Output GP (MTGP) \parencite{maddox_bayesian_2021} models, which essentially model all the outputs in parallel while additionally taking into account their correlations. The computational burden is excessive, especially for high-dimensional problems. This is why the presented methodology has been chosen over MTGP. Another point which might be addressed in the future is the use of Bayesian Neural Networks (BNN) instead of GPs for surrogate modelling. As presented in Section \ref{ch:hdbo}, the complexity of one GP depends on the number of samples $N$. Especially in high-dimensional problems, the number of samples might be very high, thus increasing the computational cost. As the authors in \textcite{snoek_scalable_2015} show, BNN does not scale with the number of samples $N$ but with the dimension $D$, thus staying constant over the whole optimisation process. This may lead to an improved efficiency. Additionally, there is a notable computational expense during hyperparameter tuning of the surrogate in the high-dimensional case. To mitigate this challenge, employing methods such as REMBO \parencite{wang_bayesian_2016} and ALEBO \parencite{letham_re-examining_2020} or (k)PCA-BO \parencite{raponi_2020,antonov_2022} presents an avenue for further reducing the computational cost. These methods operate under the assumption that certain dimensions are more significant than others, consequently reducing the number of tunable hyperparameters.




\printbibliography[heading=bibintoc,title=References]

\end{document}